\documentclass{article}
\usepackage{etoolbox}       %

\usepackage{graphicx}  %

\usepackage[parfill]{parskip}
\usepackage{amsmath,amsthm}
\usepackage{mathtools}  %
\usepackage[dvipsnames]{xcolor}         %

\newtoggle{arxiv}
\toggletrue{arxiv}

  \usepackage[parfill]{parskip}
  \usepackage[citestyle=authoryear-comp,sorting=nyt,maxbibnames=99,backend=biber]{biblatex}
  \newcommand{\citep}{\parencite}
  \newcommand{\citet}{\textcite}
  \usepackage{tcolorbox}
\tcbuselibrary{skins,breakable}

  \newlength{\defbaselineskip}
  \RequirePackage[T1]{fontenc}
  \RequirePackage[tt=false, type1=true]{libertine}
  \RequirePackage[varqu]{zi4}
  \RequirePackage[libertine]{newtxmath}

\usepackage{bm}

\usepackage[inline]{enumitem}
\usepackage{booktabs}
\usepackage[dvipsnames]{xcolor}  %
\usepackage{subcaption}
\usepackage{multirow}
\usepackage{wrapfig}
\usepackage{algorithm,algorithmicx,algpseudocode}
\usepackage{nicematrix}

\usepackage[frozencache,cachedir=.]{minted} %
\definecolor{natureteal}{RGB}{0,115,150}   % Nature's signature teal

\usepackage[colorlinks=true,linkcolor=natureteal,citecolor=natureteal,urlcolor=natureteal]{hyperref}
\usepackage[capitalise,noabbrev]{cleveref}

\usepackage{import}

\usepackage{pifont}%

    \title{From Syntax to Semantics: Geometric Stability as the Missing Axis of Perturbation Biology}
  \author{Prashant C. Raju\\{\footnotesize\texttt{rajuprashant@gmail.com}}}
  \date{}
\begin{document}

  \maketitle

\begin{abstract}
\noindent
The capacity to precisely edit genomes has outpaced our ability to predict the consequences. A cell can be genetically perfect and therapeutically useless: edited exactly as intended, yet unstable, drifting toward unintended fates, or selected for properties that compromise safety. This paradox reflects a deeper gap in how we evaluate biological intervention. Current frameworks excel at measuring what was done to a cell but remain blind to what the cell has become. We argue that this blindness stems from treating cells as collections of independent variables rather than as dynamical systems occupying positions on high-dimensional state manifolds. Drawing on Waddington's epigenetic landscape, we propose geometric stability as a missing axis of evaluation: the directional coherence of cellular responses to perturbation. This metric distinguishes interventions that guide cells coherently toward stable states from those that scatter them across the state manifold. Validation across diverse perturbation datasets reveals that geometric stability captures regulatory architecture invisible to conventional metrics, discriminating pleiotropic master regulators from lineage-specific factors without prior biological annotation. As precision medicine increasingly relies on cellular reprogramming, the question shifts from ``did the intervention occur?'' to ``is the resulting state stable?'' Geometric stability provides a framework for answering.

\end{abstract}

\section*{The Syntax-Semantics Gap}
Biology is navigating a transition analogous to the shift from assembly code to high-level programming: we have mastered the ability to write characters (DNA), but we struggle to compile stable programs (cell states). For half a century, the central dogma provided a reductionist map: DNA to RNA to Protein. In the era of observation, this map was sufficient. We sequenced, catalogued, and mapped. In the era of active authorship, where we edit genomes to cure disease and engineer cell therapies \citep{Boeke2016}, it is failing.\vspace{.15in}

The crisis manifests as a paradox: a cell can be genetically perfect and therapeutically useless. We treat genes as independent variables in a linear equation, assuming that a precise edit at a specific locus will produce a contained, predictable effect. But cells are complex adaptive systems existing on high-dimensional manifolds \citep{Wagner2016, Moon2019}. In this nonlinear regime, identical genetic perturbations can trigger divergent phenotypic trajectories depending on the cell's initial position in state space \citep{Jensen2017} or the local curvature of the regulatory landscape.\vspace{.15in}

Current evaluation metrics, designed for the linear paradigm, measure the ``syntax'' of the edit: indel rates, off-target cleavage, and sequence fidelity \citep{Brinkman2014, Tsai2014}. These metrics answer the engineer's question: ``Did I change the code?'' They fail to answer the biologist's question: ``Did the system stabilize?'' By focusing on sequence fidelity rather than state stability, we miss the structural failures that define the success or failure of a perturbation.\vspace{.15in}

Three such failures illustrate what the linear paradigm cannot see. First, selection artifacts. In human pluripotent stem cells, Cas9-induced breaks trigger p53-mediated toxicity even when targeting non-essential genes. The cells that survive are often those that have spontaneously acquired p53 mutations: a ``successful'' edit that has inadvertently selected for oncogenic potential \citep{Haapaniemi2018, Ihry2018}. Second, regulatory disruption. We now know that even ``clean'' integrations can disrupt 3D genome organization or enhancer dynamics, pushing cells toward malignancy without mutating coding sequences \citep{Cavazza2013, Lupiez2015}. On-target edits can trigger large deletions or chromothripsis invisible to standard sequencing \citep{Kosicki2018, Leibowitz2021}. Third, phenotypic heterogeneity. It is increasingly clear that two cells carrying the exact same edit often behave differently: one differentiating, one remaining stem-like. This is not noise. It reflects the initial position of each cell on the state manifold and the local geometry of that landscape \citep{Weinreb2020, Replogle2022}. \vspace{.15in}

These failures share a common feature: they occur in cell state space, not sequence space. We have mastered the syntax of the genome \citep{Doudna2014,Jiang2017,Jinek2012}. We remain blind to its semantics. To move forward, we must recognize that stability is not a property of the sequence. It is a geometric property of the state. 

\section*{The Manifold Reality}
\begin{figure}[h]
    \centering
    \includegraphics[width=\linewidth]{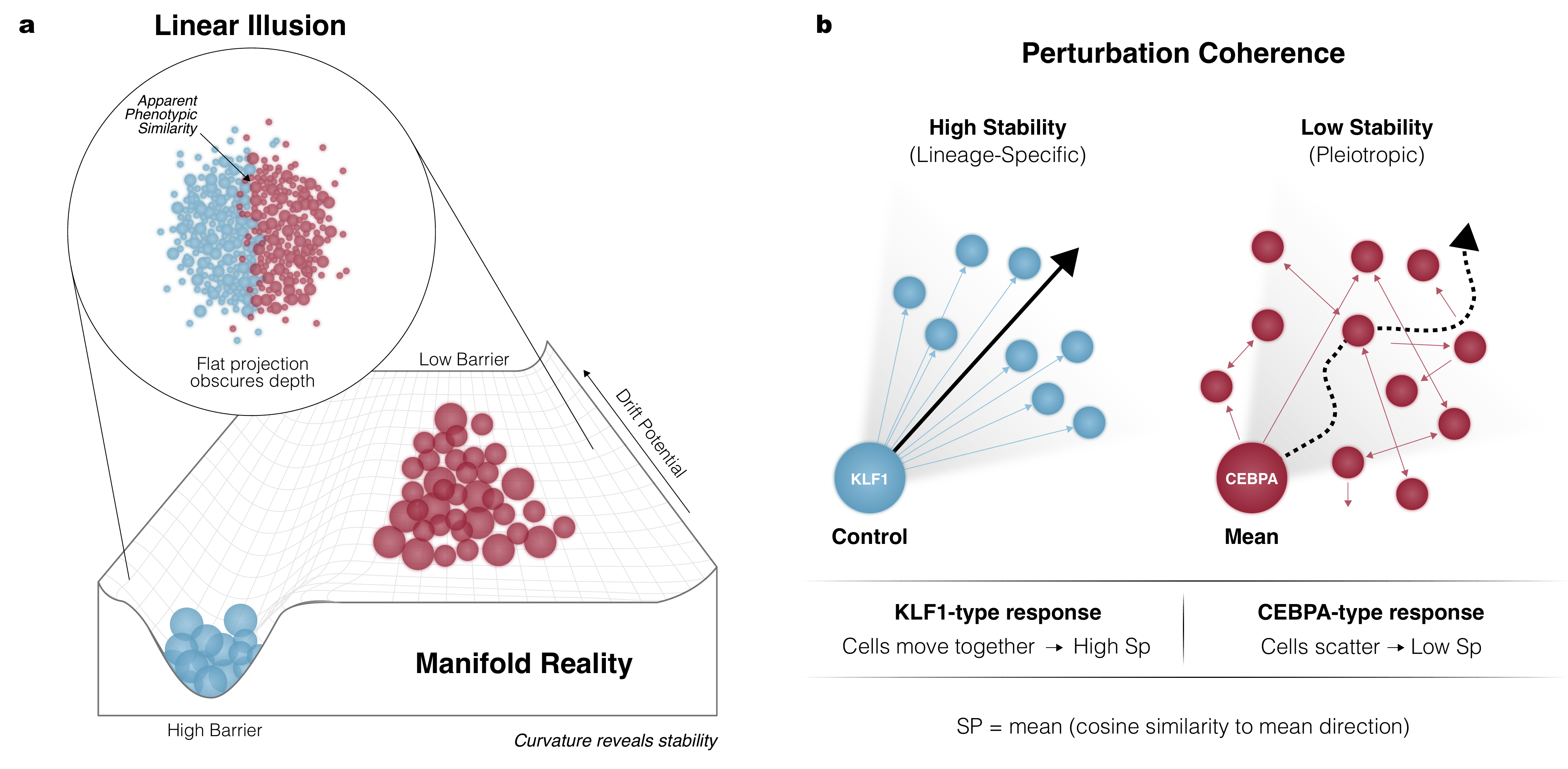}
\caption{\footnotesize\textbf{The Geometric Tax: linear metrics obscure biological stability.}
\textbf{a.}~Standard dimensionality reduction projects high-dimensional cell states onto a flat plane (Linear Illusion, inset), where two populations (blue, red) appear to overlap, suggesting similar phenotypes. Mapping these populations onto the underlying biological manifold (Manifold Reality) reveals distinct stability properties invisible to linear projections. The blue population occupies a deep valley (high barrier), representing a robust cell state resistant to perturbation. The red population sits on a shallow ridge (low barrier), representing an unstable state prone to drift. This stability difference constitutes the Geometric Tax of engineering cells into non-native configurations. \textbf{b.}~Geometric stability quantified through perturbation coherence. High-stability perturbations (left, e.g., KLF1) produce shift vectors that align coherently, indicating cells move together along a shared trajectory toward the mean direction (solid arrow). Low-stability perturbations (right, e.g., CEBPA) scatter cells in divergent directions despite similar magnitude shifts, with the mean direction (dashed arc) representing dispersed cellular responses. The Shesha stability score (Sp) captures this distinction as the mean cosine similarity between individual shift vectors and the population mean. Together, panels a and b demonstrate how manifold curvature, invisible to linear projections, determines whether perturbations produce stable or fragile cellular states.
}
\label{fig:geometric_tax}
\end{figure}

To resolve the syntax-semantics gap, we must pivot from a sequence-centric view of biology to a geometric one. This is not a radical invention but a return to the foundational insight of developmental mechanics: the epigenetic landscape. When Conrad Waddington depicted cell development as a ball rolling down an undulating surface    \citep{Waddington1957, Slack2002}, he was not merely offering an illustration  \citep{Fard2016}. He was describing the topology of a dynamical system  \citep{Ferrell2012}. In this view, ``valleys'' are not mere metaphors; they are attractor basins, stable regions of state space where regulatory networks minimize the system's quasi-potential energy  \citep{Enver2009, Huang2009, Wang2011}.\vspace{.15in}

Modern single-cell genomics has transformed this topology from a concept into a measurable reality  \citep{Rand2021}. We now understand that a human cell, expressing approximately 20,000 genes, occupies a point in a high-dimensional phase space. Gene regulatory networks function as the vector field governing this space  \citep{Raju2023b}, defining the flow that pulls cells toward attractors (stable cell types) and separates them via repellers (unstable intermediates). The fundamental law of this physics is clear: stability is geometry. A deep, steep-walled basin represents a robust state where perturbations are rapidly dampened by restoring forces, a property historically termed canalization  \citep{Siegal2002}. A shallow, flat plateau represents fragility, where transcriptional noise dominates and small fluctuations drive lineage drift  \citep{Mojtahedi2016, Elowitz2002}. While early maps of this terrain were theoretical, new computational frameworks are finally allowing us to reconstruct these landscapes directly from data  \citep{Cislo2025, Han2025}.\vspace{.15in}

Yet our analytical toolkit remains stubbornly Newtonian in an Einsteinian world. Principal Component Analysis (PCA), the workhorse of perturbation biology  \citep{Tsuyuzaki2020}, commits the ``Euclidean Error'': it forces the complex, curved geometry of the manifold onto a flat linear hyperplane  \citep{Moon2019, Zhou2021}. By measuring distances along straight lines rather than geodesic curves, we distort the very signal we seek to measure. Two cells might appear effectively identical in a 2D projection, yet be separated by a high energetic barrier that prevents state transition. In flattening the manifold, we erase the topography that defines stability. We measure the shadow of the mountain, not the mountain itself.\vspace{.15in}

Recovering the ``semantics'' of a perturbation therefore requires measuring the curvature of the manifold  \citep{Moon2019}. A cell pushed into a deep attractor moves coherently: regulatory constraints channel the perturbation vector along a defined trajectory  \citep{Huang2009}. A cell pushed onto a flat plain scatters: lacking strong restoring forces, the population diffuses stochastically  \citep{Elowitz2002, Mojtahedi2016}. This geometric divergence, coherence versus scatter, is the signature of biological meaning. It distinguishes a specific signal from entropic noise.\vspace{.15in}

\section*{Quantifying Geometric Stability}
If stability is geometry, we need metrics that measure geometry rather than magnitude. Standard perturbation analysis asks: how far did cells move in expression space? This captures effect size but misses a critical dimension  \citep{dixit2016perturb, Replogle2022}. The question that predicts functional outcomes is different: did cells move \textit{together}? A perturbation that shifts cells coherently along a shared trajectory has engaged a robust regulatory program. A perturbation that scatters cells in divergent directions has pushed them onto an unstable region of the manifold where regulatory constraints are weak  \citep{Scheffer2009}. The distinction is invisible to magnitude-based metrics but determines whether an edit produces a stable therapeutic product or a heterogeneous population primed for failure.\vspace{.15in}

The Shesha stability score operationalizes this intuition through directional coherence  \citep{raju2026geometric, raju2026canary, raju2026crispr}. Consider the hydrodynamics of a river: laminar flow moves water molecules in parallel layers, preserving local topology even as mass shifts downstream. Turbulent flow generates chaotic eddies where adjacent molecules scatter in orthogonal directions. Both regimes may transport mass the same distance, but only one preserves the structural integrity of the stream. Shesha measures whether a genetic perturbation induces laminar flow (coherent state transition) or turbulence (entropic scatter).\vspace{.15in}

A clarification is necessary. We argued above that PCA distorts global manifold geometry, and this remains true. However, what we seek to measure is not global distance but \textit{local directional coherence}: whether cells receiving the same perturbation move in similar directions relative to controls. Within the dominant principal components that capture the majority of biological variance, local angular relationships are preserved well enough to extract this signal. PCA serves as a tractable first approximation, computationally efficient and compatible with standard single-cell workflows  \citep{Tsuyuzaki2020}. The path to refinement is clear: diffusion maps, UMAP with appropriate parameterization  \citep{McInnes2018}, or learned embeddings from foundation models  \citep{Theodoris2023} offer progressively better manifold representations. But even in linear PCA space, directional coherence distinguishes biologically meaningful patterns, as validation demonstrates.\vspace{.15in}

\definecolor{grayback}{RGB}{248,248,248}
\definecolor{grayframe}{RGB}{200,200,200}

\begin{tcolorbox}[
    enhanced,
    colback=grayback,
    colframe=grayframe,
    boxrule=0.5pt,
    arc=0pt,
    outer arc=0pt,
    top=12pt,
    bottom=12pt,
    left=12pt,
    right=12pt,
    breakable
]

\textbf{\textsf{Box 1 $|$ Measuring Geometric Stability: The Shesha Framework}}

\vspace{8pt}
\hrule
\vspace{10pt}

\setlength{\parskip}{8pt}

We formalize stability as directional coherence in reduced-dimensional space. Let $\mathbf{c}$ denote the centroid of unperturbed control cells, establishing the baseline state:
\begin{equation}
\mathbf{c} = \frac{1}{n_{\text{ctrl}}} \sum_{i} \mathbf{x}_{i}^{\text{ctrl}} \tag{1}
\end{equation}
For each cell $j$ receiving perturbation $p$, the shift vector $\mathbf{v}_j = \mathbf{x}_{j}^{p} - \mathbf{c}$ captures its displacement from baseline. The population's mean shift direction $\bar{\mathbf{v}}$ and its norm define effect magnitude:
\begin{equation}
\bar{\mathbf{v}} = \frac{1}{n_p} \sum_{j} \mathbf{v}_j, \quad \text{Magnitude} = \|\bar{\mathbf{v}}\| \tag{2}
\end{equation}
Stability then follows as the average alignment between individual trajectories and the collective response:
\begin{equation}
S_p = \frac{1}{|V|} \sum_{j \in V} \frac{\mathbf{v}_j \cdot \bar{\mathbf{v}}}{\|\mathbf{v}_j\| \|\bar{\mathbf{v}}\|} \tag{3}
\end{equation}
where $V = \{j : \|\mathbf{v}_j\| > 10^{-6}\}$ excludes cells with negligible displacement. The score ranges from $-1$ to $1$: values approaching $1$ indicate \textit{laminar} response (cells moving in unison), while values near $0$ indicate \textit{turbulent} scatter (cells diffusing stochastically). Perturbations with fewer than 10 cells or mean magnitude below $10^{-6}$ are excluded to ensure statistical robustness.

\vspace{6pt}

\textbf{Implementation.} Shesha is available as a pip-installable Python package compatible with AnnData objects  \citep{Virshup2024, Virshup2023}: \texttt{pip install shesha-geometry}  \citep{shesha2026}

\end{tcolorbox}

\vspace{24pt}

Validation across four independent single-cell CRISPR datasets confirms that stability captures consistent biological signal. Analyzing 422 perturbations across 212,865 cells from CRISPRa  \citep{norman2019exploring}, CRISPRi  \citep{adamson2016multiplexed, dixit2016perturb}, and pooled screens  \citep{papalexi2021characterizing}, we observe strong magnitude-stability correlations ranging from $\rho = 0.746$ to $\rho = 0.985$, with a pooled correlation of $\rho = 0.833$ (Table~\ref{tab:correlations}). This relationship is robust across methodological choices: Euclidean distance, Mahalanobis-whitened coordinates, and local $k$-nearest-neighbor centroids all yield correlations above 0.74. The consistency across datasets, modalities, and cell types suggests a universal geometric relationship between effect size and directional coherence. Yet the correlation is not perfect: cases where magnitude and stability diverge prove biologically informative, revealing regulatory architecture invisible to magnitude alone. \vspace{.15in}

\begin{table}[H]
\centering
\caption{Magnitude-stability Spearman correlations across CRISPR datasets with 95\% bootstrap confidence intervals.}
\label{tab:correlations}
\begin{tabular}{lccccc}
\toprule
Dataset & Perturbations & Cells & $\rho_{\text{Mag}}$ & 95\% CI & $p$ \\
\midrule
Norman 2019 & 236 & 99,420 & 0.953 & [0.934, 0.965] & $< 10^{-100}$ \\
Adamson 2016 & 8 & 5,752 & 0.929 & [0.407, 1.000] & $< 10^{-3}$ \\
Dixit 2016 & 153 & 89,350 & 0.746 & [0.641, 0.827] & $< 10^{-100}$ \\
Papalexi 2021 & 25 & 18,343 & 0.985 & [0.939, 0.997] & $< 10^{-18}$ \\
\midrule
Pooled & 422 & 212,865 & 0.833 & --- & $< 10^{-100}$ \\
\bottomrule
\end{tabular}
\end{table}

\begin{figure}[H]
\centering
\includegraphics[width=\textwidth]{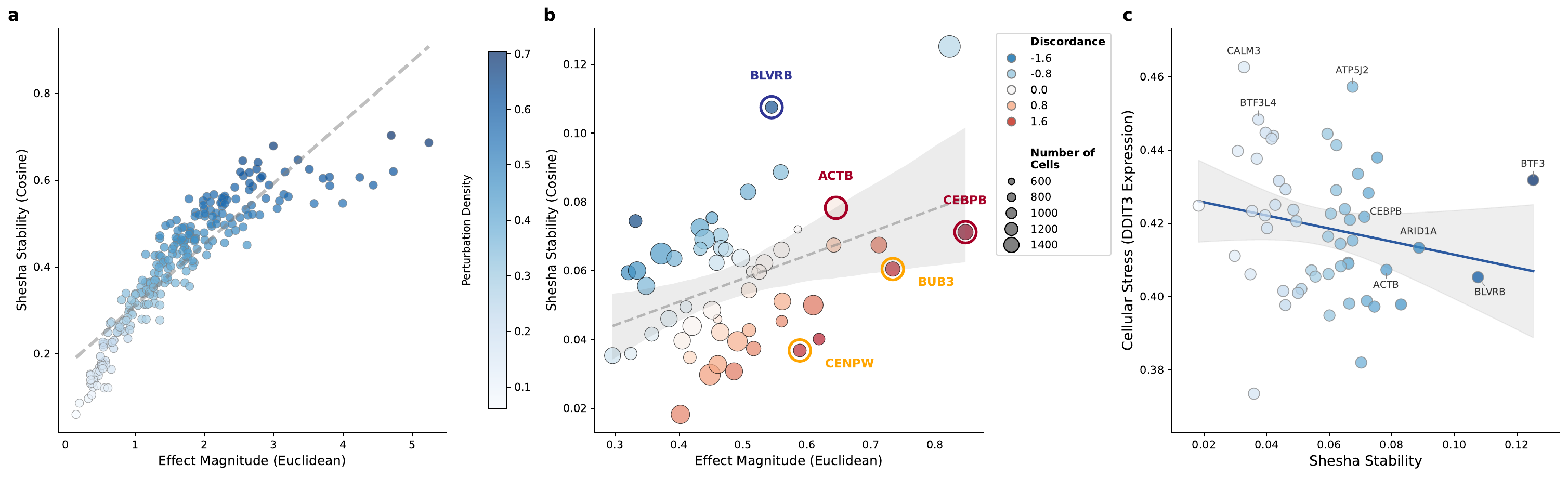}
\caption{\footnotesize\textbf{Geometric stability validated across CRISPR datasets and linked to cellular stress.}
\textbf{a.}~Magnitude-stability relationship in Norman et al.\ CRISPRa dataset   \citep{norman2019exploring} ($n=236$ perturbations). Shesha stability score correlates strongly with effect magnitude (Spearman $\rho=0.953$, $p<10^{-100}$). Color indicates local perturbation density. Dashed line shows linear fit.
\textbf{b.}~Independent validation in the Replogle et al.\ genome-scale CRISPRi screen   \citep{Replogle2022} (K562 cells). Point color indicates discordance (deviation from expected stability given magnitude); point size indicates cell count per perturbation. Labeled genes illustrate biological interpretation: BLVRB (biliverdin reductase, metabolic) shows high stability relative to magnitude, consistent with pathway-specific effects. CEBPB (C/EBP family transcription factor) and CENPW (centromere protein) show low stability relative to magnitude, consistent with pleiotropic or cell-division-wide effects. BUB3 (spindle checkpoint) demonstrates that low stability is not merely a proxy for cell cycle arrest. Grey shading indicates 95\% confidence interval.
\textbf{c.}~Functional consequence of geometric instability. Stability negatively correlates with DDIT3 expression (Spearman $\rho=-0.28$, $p=0.041$), a canonical marker of cellular stress  \citep{Oyadomari2003}. Perturbations producing incoherent cellular responses (low stability) induce elevated stress signatures. Point shading encodes stability (darker = higher). Linear fit shown with 95\% confidence interval. Notably, no perturbations occupy the high-stability/high-stress quadrant, suggesting that geometric coherence is a prerequisite for cellular homeostasis.}
\label{fig:validation}
\end{figure}

Critically, however, the magnitude-stability trade-off is not an artifact of linear projection. Validation using embeddings from scGPT, a foundation model trained on 33 million cells that learns nonlinear representations of cell state  \citep{Cui2024}, confirms that this geometric relationship persists robustly in nonlinear latent space ($\rho = 0.935$, 95\% CI $[0.911, 0.951]$, $p < 10^{-107}$ vs. $\rho = 0.953$ in PCA for Norman et al. 2019  \citep{norman2019exploring}). While diffusion maps  \citep{Coifman2006}, PHATE  \citep{Moon2019}, and other manifold-aware embeddings may reveal additional structure, this indicates that the tax is a property of biological state space itself, independent of dimensionality reduction technique.\vspace{.15in}

\section*{Discordance Reveals Regulatory Architecture}
The strong correlation between perturbation magnitude and geometric stability might suggest these metrics are redundant. They are not. The divergent cases, where magnitude and stability decouple, expose a fundamental distinction in how genes shape the regulatory landscape. This divergence captures the difference between regulators that coordinate cellular programs and those that merely perturb them.\vspace{.15in}

Master regulators pay a Geometric Tax. CEBPA, the canonical driver of myeloid differentiation, activates over 24 downstream pathways spanning immune response, cell cycle control, metabolism, and lineage commitment   \citep{Friedman2007, Friedman2015, norman2019exploring}. When CEBPA is perturbed, cells move far from controls in transcriptomic space, but they scatter incoherently, each cell activating a different subset of CEBPA's broad target repertoire. This is not a failure of the perturbation; it is a consequence of regulatory architecture. A single node attempting to coordinate thousands of competing variables cannot maintain geometric coherence. The Tax is the cost of pleiotropy: large effects, diffuse outcomes. Despite ranking among the highest-magnitude perturbations in the Norman CRISPRa dataset, CEBPA combinations cluster below the regression line, their stability lower than effect size would predict.\vspace{.15in}

Lineage-specific factors operate under different dynamical regime. KLF1, the erythroid-specific transcription factor essential for globin expression and terminal red blood cell maturation  \citep{Miller1993, Tallack2010_GR, Tallack2010_IUBMB, Siatecka2011}, regulates a narrower but tightly coordinated gene program. KLF1 perturbations produce moderate magnitude shifts but exceptional coherence: cells move together along a shared trajectory corresponding to erythroid differentiation arrest  \citep{Pilon2008}. KLF1 pays minimal Geometric Tax because its targets are functionally aligned, part of a single regulatory module rather than distributed across the genome. The stability score captures this alignment without any prior knowledge of KLF1's biology.\vspace{.15in}

This distinction, pleiotropic master regulators versus lineage-specific factors, generalizes across datasets and screening modalities. In an independent genome-scale CRISPRi screen with over 2,000 perturbations in K562 cells  \citep{Replogle2022}, CEBPB, a member of the same transcription factor family as CEBPA  \citep{Jakobsen2013}, shows the identical signature: high magnitude, low stability. CENPW, encoding a centromere protein required for chromosome segregation  \citep{Hori2008, Liu2024}, shows even more pronounced discordance. Disrupting the cell division machinery produces large but geometrically incoherent responses as cells scatter into diverse failure states. Conversely, BLVRB shows high stability despite moderate magnitude, consistent with its role encoding biliverdin reductase, an enzyme in a specific metabolic pathway  \citep{OBrien2015, Paukovich2018}. The pattern is not dataset-specific. It reflects how regulatory architecture shapes the topology of state space.\vspace{.15in}

The statistical signature confirms what the landscape framework predicts. Perturbations in the high-discordance quartile (high magnitude, low stability) exhibit nearly three-fold higher variance in stability scores ($SD = 0.202$) compared to low-discordance perturbations ($SD = 0.074$). Shallow valleys produce scatter; deep valleys produce coherence. The variance itself encodes the geometry.\vspace{.15in}

The Geometric Tax has functional consequences beyond phenotypic heterogeneity. Across essential gene perturbations in the Replogle dataset, stability negatively correlates with DDIT3 expression (Spearman $\rho = -0.28$, $p = 0.041$), a canonical marker of the integrated stress response  \citep{Zinszner1998, Oyadomari2003}. Perturbations that scatter cells incoherently across transcriptomic space induce elevated stress signatures. Geometric instability is not merely a statistical abstraction; it manifests as cellular dysfunction.\vspace{.15in}

One might argue that stability simply measures proliferation arrest: cells that stop dividing would appear artificially coherent. The data refute this. CENPW  \citep{Zhang2024} and BUB3  \citep{Logarinho2008}, both cell cycle regulators whose perturbation induces mitotic arrest, show low stability despite halting division. BLVRB  \citep{OBrien2015} shows high stability while cells continue cycling. The metric captures regulatory coherence, not cessation of proliferation.\vspace{.15in}

The implications extend beyond biological understanding. For CRISPR screens, geometric stability can prioritize hits likely to replicate, distinguishing perturbations that engage specific programs from those causing nonspecific toxicity  \citep{Tycko2019}. For cell therapy manufacturing, stability can flag products at risk of lineage drift or functional heterogeneity before clinical deployment  \citep{Lipsitz2016, Kliegman2024}. Indel rates answer whether the edit occurred  \citep{Sentmanat2018}. Stability answers whether the resulting state will persist.\vspace{.15in}

\section*{Stability as a Design Principle}
The preceding sections establish geometric stability as a measurable property of perturbations. But stability is more than a metric. It is a design principle that evolution discovered long before we learned to measure it.\vspace{.15in}

Natural selection operates on phenotypes, not genotypes  \citep{Mayr1963}. A mutation that produces large transcriptional effects but scatters cells into incoherent states will be rapidly purged: such cells cannot coordinate the programs required for survival, proliferation, or function  \citep{Fisher1930, Kirschner1998}. Conversely, mutations that engage robust regulatory modules, moving cells coherently along trajectories that preserve functional capacity, can propagate  \citep{Kirschner1998, Raju2023}. Over evolutionary time, this selection pressure has carved the deep valleys of Waddington's landscape  \citep{Waddington1957, Wang2011}. The attractor basins we observe in single-cell data are not arbitrary; they are the stable configurations that have survived billions of years of geometric selection  \citep{Huang2009, Wagner2016, Raju2023b}. While early theoretical work showed that such canalization emerges naturally in evolving networks  \citep{Siegal2002}, recent studies have mathematically formalized Waddington's landscape, demonstrating how these geometric constraints direct evolutionary trajectories  \citep{Raju2023}. Thus, gene regulatory networks are optimized not merely for specific expression patterns but for the stability of those patterns under perturbation.\vspace{.15in}

This framing recasts the challenge of biological engineering. When we edit a genome, we are not simply changing a sequence; we are perturbing a system that evolution has optimized for stability  \citep{Kitano2004}. The Geometric Tax is the cost of fighting this optimization. Edits that work with the grain of regulatory architecture, engaging existing modules and respecting attractor boundaries, pay minimal tax. Edits that work against it, forcing cells into states that evolution never selected for, pay heavily in the currency of heterogeneity, stress  \citep{Zinszner1998}, and drift  \citep{Weinreb2020}.\vspace{.15in}

This is biology's alignment problem. In artificial intelligence, alignment refers to the challenge of ensuring that optimized systems pursue intended objectives rather than exploiting loopholes in their reward functions  \citep{Amodei2016, Krakovna2020, Hubinger2019, Russell2019}. In biological engineering, the analogous challenge is ensuring that edited cells maintain intended phenotypes rather than drifting toward unintended attractors. A CAR-T cell optimized for tumor killing may instead find a local minimum in exhaustion  \citep{Fraietta2018, Philip2017, Sen2016}. An iPSC-derived beta cell may appear differentiated by marker expression yet occupy a shallow basin that permits dedifferentiation under metabolic stress  \citep{Lipsitz2016, Nair2019, Talchai2012}. Current evaluation frameworks, focused on sequence fidelity and marker expression, cannot detect misalignment until it manifests clinically  \citep{Bravery2013, Capelli2023, Fraietta2018, Salmikangas2015}. Geometric stability offers an objective function for alignment: perturbations that produce high stability are those where the intended state coincides with a deep attractor.\vspace{.15in}

The practical implications follow from this principle. In screening, stability prioritizes hits that engage robust regulatory programs over those that produce large but incoherent effects, predicting which perturbations will replicate across contexts  \citep{Tycko2019}. In manufacturing, stability flags cell products balanced on flat manifold regions before clinical deployment, complementing marker-based QC with a measure of phenotypic robustness  \citep{Levine2017, Lipsitz2016, Fraietta2018}. In regulatory review, stability addresses a failure mode that current genotoxicity and tumorigenicity assays cannot access: the edited cell that is genetically clean but phenotypically fragile  \citep{FDA2013, FDA2024}.\vspace{.15in}

This imperative extends beyond the bench to a transformation underway in computational biology. Foundation models trained on transcriptomic data, including scGPT  \citep{Cui2024}, Geneformer  \citep{Theodoris2023}, and Universal Cell Embeddings  \citep{Rosen2023}, learn representations where distances correspond to biological relationships. These models implicitly encode the geometry of cell state space and inherit the same fragility problem. Geometric stability operationalizes a question that matters equally for evaluating any learned biological embedding: do perturbations move cells coherently through latent space, or do they scatter them? A foundation model that places biologically stable states in deep basins and unstable intermediates on flat regions has learned something true about cellular dynamics. A model that scrambles this geometry has learned a representation that will fail when deployed. Geometric stability thus serves as a necessary criterion for representation quality.\vspace{.15in}

The deeper point transcends any particular implementation. Evolution has spent billions of years optimizing gene regulatory networks for stability. We are only now developing the tools to measure what it optimized for. As biological engineering matures from sequence writing to state programming, geometric stability offers a bridge: between the intuitions of machine learning and the constraints of cellular dynamics, between what we intend and what cells become.\vspace{.15in}

% Looking forward, geometric stability addresses a critical challenge for the emerging era of generative biology. As the field transitions toward multimodal foundation models capable of \textit{in silico} perturbation   \citep{Cui2025}, these systems require a physical constraint to distinguish viable biological attractors from hallucinatory intermediate states  \citep{Rathkopf2025}. Shesha provides this validation framework by quantifying whether predicted states occupy stable manifold regions, ensuring that computational predictions remain faithful to biological dynamics.\vspace{.15in}

% Finally, this framework provides necessary quality control for synthetic data. As the field adopts foundation models for \textit{in silico} perturbation   \citep{Cui2025}, we face a new validation gap: distinguishing viable biological attractors from hallucinatory intermediates  \citep{Rathkopf2025}. Just as Shesha identifies unstable cells in a Petri dish, it serves as an objective function for generative models, ensuring that computational predictions remain faithful to biological dynamics.\vspace{.15in}

Finally, this framework resolves a critical challenge for the emerging era of generative biology. As foundation models gain the capability for \textit{in silico} perturbation  \citep{Cui2025}, we face a new validation gap: distinguishing viable biological attractors from hallucinatory intermediates  \citep{Rathkopf2025}. Just as Shesha identifies unstable cells in experimental screens, it provides the necessary physical constraint for generative models, ensuring that computational predictions remain faithful to biological dynamics. Geometric stability becomes the physics engine that keeps our digital explorations grounded in cellular reality.\vspace{.15in}

% Looking forward, this metric resolves a critical challenge for the emerging era of generative biology. As the field transitions toward multimodal foundation models capable of \textit{in silico} perturbation  \citep{Cui2025}, geometric stability offers the necessary physical constraint. While these models can predict state transitions, they require a stability metric to distinguish viable biological attractors from hallucinatory latent states  \citep{Rathkopf2025}. Shesha provides this `physics engine' for the generative era, ensuring that our digital maps remain faithful to the biological territory.\vspace{.15in}

\section*{What the Thunder Said}

We have argued that the central challenge of biological engineering is not writing the genome but compiling stable programs from it. The syntax-semantics gap is not a limitation of current tools; it is a consequence of applying linear intuitions to nonlinear systems. Cells are not collections of independent variables. They are dynamical systems occupying positions on high-dimensional manifolds, governed by vector fields that evolution has spent billions of years optimizing. When we edit a genome, we perturb this system. The outcome depends not on the edit alone, but on the geometry of the landscape at the point of perturbation.\vspace{.15in}

This geometry imposes a tax. Master regulators like CEBPA, coordinating dozens of downstream pathways, pay heavily: their perturbations produce large effects but scatter cells into incoherent states. Lineage-specific factors like KLF1, engaging tightly coordinated modules, pay minimally: their perturbations move cells together along defined trajectories. The Geometric Tax is not a metaphor. It is the measurable cost of working against regulatory architecture, quantifiable as the divergence between effect magnitude and directional coherence. Current evaluation frameworks, designed for the linear paradigm, cannot see this cost. They measure the shadow of the manifold, not the manifold itself.\vspace{.15in}

Geometric stability fills this gap. It asks not just whether the edit occurred, but whether the resulting state is robust. It distinguishes laminar response from turbulent scatter, deep attractors from shallow plateaus, aligned perturbations from misaligned ones. Validation across CRISPR modalities demonstrates that this measure captures regulatory architecture invisible to sequence-level metrics. The CEBPA-KLF1 distinction emerges from the data without supervision, as does the correlation between instability and cellular stress. The geometry is real. It can be measured. And it predicts functional outcomes that magnitude alone cannot.\vspace{.15in}

As biological engineering matures from sequence editing to state programming, evaluation must follow. Efficiency, specificity, and stability should become the three axes of assessment: efficiency to confirm the edit occurred; specificity to confirm it occurred only where intended; and stability to confirm the resulting state will persist. The first two axes are well established. The third is the geometric imperative.\vspace{.15in}

We have learned to spell the genome. What remains is to learn its grammar. In the wasteland of unstable phenotypes, where raw edits bring only noise, the thunder finally speaks  \citep{eliot1922wasteland}. Its command is simple: Control the geometry.\vspace{.15in}

\section*{Author Competing Interests}
The author declares no competing interests.

\section*{Declaration of generative Al and Al-assisted technologies in the writing process}
During the preparation of this work the author used Claude Sonnet 4.5, Claude Opus 4.5, and Gemini 3 Pro in order to improve the quality of the writing. After using this tool, the author reviewed and edited the content as needed and takes full responsibility for the content of the published article.

% \section{Data Availability}
% The five publicly available single-cell transcriptomic datasets analyzed in this study were accessed through the pertpy framework\citep{pertpy}. These include the genome-scale Perturb-seq maps from Replogle et al. (2022) (DOI: 10.1016/j.cell.2022.05.013), and the perturbation screens from Norman et al. (2019) (GSE133344), Adamson et al. (2016) (GSE90546), Dixit et al. (2016) (GSE90063), and Papalexi et al. (2021) (DOI: 10.1038/s41588-021-00778-2). No new biological data were generated for this work.\vspace{.15in}

\section*{Code Availability}
The full code necessary to reproduce all experiments, benchmarks, and analysis described in this paper is publicly available at \url{https://github.com/prashantcraju/geometric-stability-crispr}. 

\section*{Acknowledgments}
We thank Padma K. and Annapoorna Raju for generously supporting the computational resources used in this work. We thank the authors of the datasets that were used for making their data publicly available.

\printbibliography

\end{document}